\documentclass[twocolumn,showpacs,preprintnumbers,amsmath,amssymb]{revtex4}
\usepackage{graphicx}
\begin{document}

\title{One-body Properties of Nuclear Matter with Off-shell Propagation}
\baselineskip=1. \baselineskip

\author{P. Bo\.{z}ek}
\email{piotr.bozek@ifj.edu.pl}
\affiliation{Institute of Nuclear Physics, PL-31-342 Cracow, Poland}

\date{\today}





\begin{abstract}

Symmetric nuclear matter is studied in the self-consistent,
in-medium $T$-matrix approach. One-body spectral function, optical
potential, and scattering width are calculated. Properties of quasi-particle
excitations at the Fermi surface are discussed. Dispersive self-energies are
dominated by contributions from the $^1S_0$, $3S_1-^3D_1$, and $^3P_1$
partial waves.
\end{abstract}

\keywords{ 
Nuclear matter,
 optical potential, Fermi-liquid properties}


\pacs{21.65+f, 24.10Cn}


\maketitle

\section{Introduction}

Single-particle properties of nucleons are modified inside the strongly
interacting nuclear matter\cite{Mahaux}. The optical potential describes
 the average interaction
of a nucleon with the medium. 
 At negative energies, the single-particle potential
reflects the modifications of the quasi-particles in the nuclear matter.
The spectral function of a nucleon is a measure of the energy distribution of
a plane wave state in the system, and can be observed in electron scattering 
experiments.

Single-particle properties can be calculated from the 
Brueckner-Hartree-Fock (BHF) theory \cite{kohler92,baldoetal92}. 
Hole-hole diagrams must be added to the lowest order results 
to obtain  a meaningful imaginary part of the self-energy.
Because of the violation of the Hugenholz-Van Hove property by the BHF 
approach, the overall normalization of the single-particle energies is
sometimes a problem. BHF calculations use a quasi-particle approximation
for the propagators; therefore, a finite order diagram has 
kinematical limits on the total energy in the scattering.
A finite total energy in the interaction could introduce distortions
in the far energy tails of the calculated spectral functions.

Recently the in-medium $T$-matrix approximation has been applied to the
nuclear matter
\cite{Vonderfecht:1996,Alm:1996ps,di1,Bozek:1998su,Dickhoff:1999yi,Dewulf:2000jg,Bozek:2001tz}.
One obtains nontrivial self-energies by summing a ladder of hole-hole and 
particle-particle propagators. The self-energies obtained are
$\Phi$-derivable \cite{Baym,Bozek:2001tz}; within this approximation, the
single-particle properties are consistent  with global thermodynamical
properties of the system. A necessary requirement for the in-medium 
$T$-matrix approach is to use fully self-consistent self-energies and
propagators. The propagators are dressed with the imaginary and real parts of
the self-energy, and the full spectral function for nucleons must be taken.
It requires  a serious numerical effort in the calculations, but
using a quasi-particle approximation in the $T$-matrix ladder 
gives too strong pairing \cite{Bozek:1998su,Bozek:2000fn,Dickhoff:1999yi},
too much scattering \cite{Dickhoff:1999yi,Bozek:1998su},
and incorrect single-particle energies \cite{Bozek:2001tz}.

In the present work we calculate single-particle self-energies and spectral 
functions using fully self-consistent propagators. We take  the complete energy
dependence of the self-energies and spectral functions into account in order
access reliably  the high energy tails of the spectral functions.
We present results on the single-particle potential,
scattering width and spectral function, obtained with a separable  
Paris interaction with several partial waves 
(App. \ref{NNint}). In Appendix \ref{nummeth} we describe in details 
the numerical methods that allow us  to tackle the problem.

We find that    normal nuclear density is at the limit of the 
region of the $^3S_1-^3D_1$ pairing instability. Owing to the
dressing of propagators in the gap equation the superfluid gap 
is very small, if not vanishing 
(depending on the details of the interaction in the deuteron
channel).
 We leave the detailed analysis of the pairing
in symmetric nuclear matter with realistic interactions to a different work,
and study in the following the normal state of nuclear matter.

\section{In-medium $T$-matrix}

In this Section we recall the basic equations of the approach used,
more details and discussion can be found in 
\cite{KadanoffBaym,Danielewicz:1984ca,Kraeft,Alm:1996ps,
Vonderfecht:1996,di1,Bozek:1998su}.
The $T$-matrix approximation sums ladder diagrams with dressed 
particle-particle and hole-hole propagators for the in-medium two-particle 
propagator. The retarded $T$-matrix is
\begin{eqnarray}
\label{teq}& &
\langle{\bf p}|T({\bf P},\Omega)|{\bf p}^{'}\rangle 
  = V({\bf p},{\bf p}^{'}) 
\nonumber \\  & & + 
 \int\frac{d\omega_1d\omega_2d^3q}{32\pi^5}
 V({\bf p},{\bf
  q}) A(p_1,\omega_{1})A(p_2,\omega_{2})\nonumber \\ & & 
\frac{\big(1-f(\omega_1)-f(\omega_2)\big)}
{\Omega-\omega_1-\omega_2+i\epsilon}
 \langle{\bf q}|T({\bf P},\Omega)
|{\bf p}^{'}\rangle \end{eqnarray}
where ${\bf p_{1,2}}={\bf P}/2\pm {\bf q}$~ and $f(\omega)$ is the Fermi 
distribution.
A partial wave expansion of the in-medium $T$-matrix is performed
\begin{eqnarray}
\label{teqpw}& &
\langle p|T^{(JST)}_{l l^{'}}( P,\Omega)|p^{'}\rangle 
  = V^{(JST)}_{l l^{'}}({ p},{ p}^{'}) 
\nonumber \\  & & + \sum_{l^{''}}
 \int\frac{d\omega_1d\omega_2q^2dq}{8\pi^4}
 V^{(JST)}_{l l^{''}}({ p},{
  q})\nonumber \\
& & \langle A(p_1,\omega_{1})A(p_2,\omega_{2})\rangle_{P,q}\nonumber \\ & & 
\frac{\big(1-f(\omega_1)-f(\omega_2)\big)}
{\Omega-\omega_1-\omega_2+i\epsilon}
 \langle{ q}|T^{(JST)}_{l^{''} l^{'}}({ P},\Omega)
|{ p}^{'}\rangle 
\end{eqnarray}
after angle averaging the intermediate two-particle propagator
($\langle \dots \rangle =\int\frac{d\Omega}{4\pi} \dots$)

The imaginary part of the retarded self-energy is obtained by
closing a pair of external vertices in the $T$-matrix with a fermion 
propagator
\begin{eqnarray}
\label{imags}
& & 
{\rm Im}
\Sigma(p,\omega) =\int\frac{d\omega_1d^3k}{16 \pi^4}
A(k,\omega_1) \nonumber \\ 
& & \langle({\bf p}-{\bf k})/2|{\rm Im}T({\bf p}
+{\bf k},\omega+\omega_1)|({\bf p}-{\bf k})/2\rangle_A \nonumber \\& &
 \Big( f(\omega_1)+b(\omega+\omega_1) \Big) 
\end{eqnarray}
where 
\begin{eqnarray}
\label{spectralf} 
A(p,\omega)= ~~~~~~~~~~~~~~~~~~~~~~~~~& & \nonumber\\  \frac{-2 \mathrm{Im}
\Sigma(p,\omega)}{\left(\omega-p^2/2m 
-\mathrm{Re}\Sigma(p,\omega)\right)^2 +\mathrm{Im}\Sigma(p,\omega)^2} & &
\end{eqnarray}
is the self-consistent spectral function of the nucleon and $b(\omega)$
is the Bose distribution.
The real part of the self-energy is related to ${\rm Im} \Sigma$
by a dispersion relation
\begin{equation}
\label{reals}
\mathrm{Re}\Sigma(p,\omega)= \Sigma_{HF}(p) + {\cal P} \int \frac{d\omega^{'}}
{\pi} \frac{\mathrm{Im}\Sigma(p,\omega^{'})}{\omega^{'}-\omega}
\end{equation}
with $\Sigma_{HF}(p)$ being the Hartree-Fock self-energy.
Eqs. (\ref{teq}), (\ref{imags}), (\ref{reals}) and (\ref{spectralf}) 
are to be solved iteratively
and at each iteration the chemical potential $\mu$ is adjusted to fulfill
the condition on the density 
\begin{equation}
\label{densitycon}
\rho=\int\frac{d\omega d^3p}{16\pi^4}
A(p,\omega)f(\omega)=.16\textrm{fm}^{-3} \ .
\end{equation}

The $T$-matrix approximation sums ladder diagrams contributing to the 
ground state energy. In this way it regularizes the short range core in the
nucleon interaction similarly as the BHF approach. At the same time, the
self-consistent $T$-matrix approximation is $\Phi$-derivable \cite{Baym}.
It was shown for a model interaction that if a fully self-consistent
$T$-matrix calculation is performed the thermodynamic consistency relations
are fulfilled \cite{Bozek:2001tz}.
In the $T$-matrix approximation a nonzero imaginary self-energy appears and
leads to a nontrivial spectral function. It is crucial to keep the full
off-shell dependence of the propagators in the calculations. It is possible to
treat the resulting energy integrals using  numerical methods
described in the Appendix \ref{nummeth}.
All the calculations are done using the Fermi energy as the origin of the
energy scale. The figures are also plotted using this convention.
The calculations are done using a separable parameterization of the Paris
potential  \cite{parisseparable,parisseparable2}.
The Fermi energy obtained is $-21.4\textrm{MeV}$ and the binding 
energy $-15.1\textrm{MeV}$ at normal density. This is not the  saturation
point for this interaction, but the values quoted above are quite reasonable
and give some confidence in the single-particle properties we want to study.

\section{Optical potential}

The real-part of the self-energy defines 
the single-particle pole in the propagator
$$
\omega_p=p^2/2m+\textrm{Re}\Sigma(p,\omega_p) \ .
$$
The free dispersion relation is modified due to interactions 
with the medium. The resulting 
effective potential $\textrm{Re}\Sigma(p,\omega_p)$
is attractive. It leads in particular to a reduction  
of the effective mass $m^\star=\frac{pdp}{d\omega_p}$. We find $m^\star\simeq
.85m$ at the Fermi momentum, this value depends somewhat on
the chosen interaction. The effective mass can be written \cite{Mahaux}
 as the product of
the $k$-mass $m_k/m=\left(1+\frac{m\partial\textrm{Im}\Sigma(k,\omega)}
{k\partial k}|_{\omega=\omega_k}\right)^{-1}$ and the $\omega$-mass
$m_\omega/m=\left(1-\frac{\partial\textrm{Im}\Sigma(k,\omega)}
{\partial \omega}|_{\omega=\omega_k}\right)$. We find
$m_k/m\simeq.62$ and $m_\omega/m\simeq1.37$ at the Fermi momentum.

\begin{figure}
\includegraphics[width=0.48\textwidth]{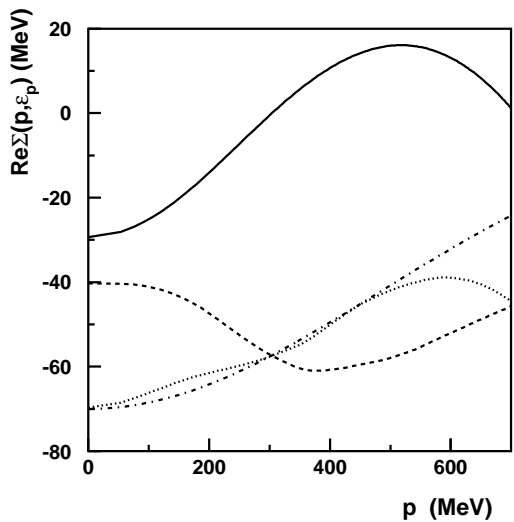}
\caption{\label{realpot} The 
real part of the self-energy on-shell (solid line),
the Hartree-Fock self-energy (dotted line), and the dispersive part of the
self-energy (dashed line).}
\end{figure}

The real part of the self-energy is the sum of the Hartree-Fock and
the dispersive contribution (\ref{reals}). The 
 real part of the dispersive self-energy is negative around the Fermi energy
and around the quasi-particle pole (Fig. \ref{realpot}). 
The value of the real part of the self-energy on-shell is the effective
potential felt by the quasi-particle in the medium. For positive energies of
the particle it corresponds to the optical potential.
The depth of the single particle potential is around $-70\textrm{MeV}$
and it is decreasing with momentum. In the range of positive single particle
energies ($280\textrm{MeV}<p<500\textrm{MeV}$), the single-particle potential
 can be fitted 
with the form \cite{perey}
$$
\label{fitopt}
\textrm{Re}\Sigma(p,\omega_p)\simeq H_0\exp\left(-\beta^2k^2/4\right)
$$
with
$H_0=-70\textrm{MeV}$ and $\beta=.58\textrm{fm}$ (dashed-dotted line in Fig.
\ref{realpot}).
The range of nonlocality $\beta$ is small reflecting the fact that the
effective mass $m^\star$ is not very different from the free one.
The depth of the potential for positive energies is consistent with values
obtained from 
phenomenological analyzes of the optical potential \cite{optpot}.
The single particle potential is weakly dependent on the
temperature; at $T=4\textrm{MeV}$ $\textrm{Re}\Sigma(p,\omega_p)$
is shifted up by less than $1\textrm{MeV}$ from its value at zero temperature.
\begin{figure}
\includegraphics[width=0.48\textwidth]{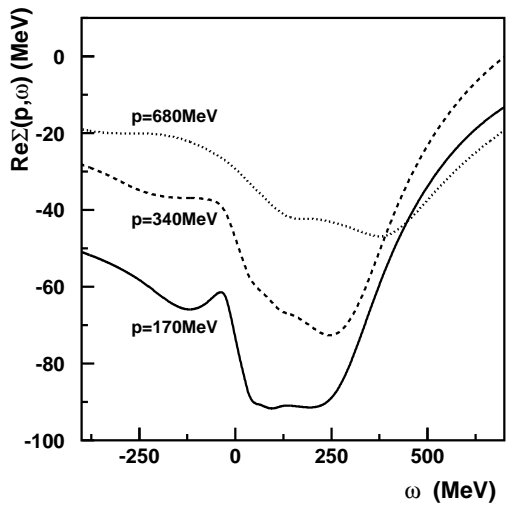}
\caption{\label{realofig} The energy dependence of the single particle potential 
$\textrm{Re}\Sigma(p,\omega)$ at fixed momentum.}
\end{figure}

In Fig. \ref{realofig} is presented the energy dependence of the real part of
the self-energy. The energy dependence of the single particle potential is
responsible for the time nonlocality of the optical potential. 
In the whole range of frequencies in Fig. \ref{realofig} some energy
dependence of the potential can be seen. This energy dependence is decreasing
with momentum. For energies close to the Fermi energy the energy dependence
of the single-particle potential is similar to the one presented 
in Ref. \cite{baldoetal92}. The energy dependence 
determines the quasi-particle strength
\begin{equation}
\label{zfac}
Z_p= \left( 1 - \frac{\partial
    \textrm{Re}\Sigma(p,\omega)}{\partial\omega}|_{\omega=\omega_p}\right) \ .
\end{equation}
 At the Fermi momentum we find $Z_{p_F}\simeq.73$.
On the other hand, in the range 
$50\textrm{MeV}<\omega<200\textrm{MeV}$ $\textrm{Re}\Sigma(p,\omega)$ is
relatively flat as function of energy.

\section{Scattering width}

\begin{figure}
\includegraphics[width=0.48\textwidth]{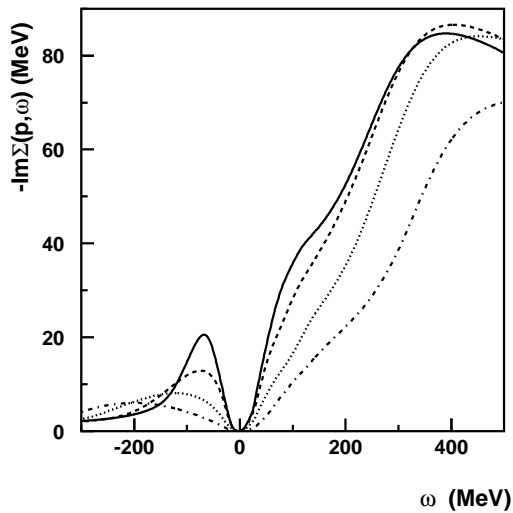}
\caption{\label{imagfig} The single particle width  
$-\textrm{Im}\Sigma(p,\omega)$ as function
  of energy for $p=0$, $170$, $340$, and $510\textrm{MeV}$; solid, dashed,
  dotted, and dashed-dotted lines, respectively.}
\end{figure}
The off-shell scattering width is nonzero on both sides of the Fermi energy 
but
larger at positive energies (Fig. \ref{imagfig}).
 At negative energies the tail of the scattering
width extends very far. It is a general 
feature of self-consistent calculations.
Dressed propagators in the $T$-matrix ladder make the imaginary part of the
$T$-matrix nonzero even at very negative energies. On the other hand, 
quasi-particle approximations have a kinematical limit on the lowest energy
in the scattering. The scattering width decreases with momentum. It can be
understood by the fact
that at low momenta the interaction is dominated by the strongest $S$ waves.
Other calculations \cite{fabrocini,kohler92,baldoetal92,Bozek:1999se,lehr,Bozek:1998su,Dickhoff:1999yi,Dewulf:2000jg} 
of the
variational, BHF, 
 Born, or $T$-matrix type give qualitatively but not quantitatively 
similar results.
The differences are partly due to
different short range properties of the interactions used. 

Consistently with
 general features of Fermi liquids  at zero temperature, we find that 
 $\textrm{Im}\Sigma(p,\omega=0)=0$ (Fig. \ref{imagfig}). The same relation, 
coming form the
  restricted phase-space, is fulfilled by other approximations
\cite{fabrocini,baldoetal92,lehr}. 
At higher temperatures a nonzero scattering width at the Fermi energy 
appears.
\begin{figure}
\includegraphics[width=0.48\textwidth]{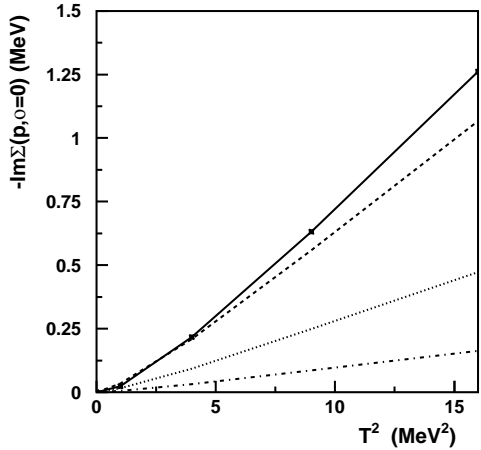}
\caption{\label{imagtemp} The temperature dependence of the single particle width
  $-\textrm{Im}\Sigma(p,\omega=0)$ at the Fermi energy
  for $p=0$, $170$, $340$, and $510\textrm{MeV}$; solid, dashed,
  dotted, and dashed-dotted lines, respectively.}
\end{figure}
As can be seen from Fig. \ref{imagtemp} the
scattering width at the Fermi energy is increasing as the square of the
temperature, as expected from phase space arguments \cite{abrikosov}.

\begin{figure}
\includegraphics[width=0.48\textwidth]{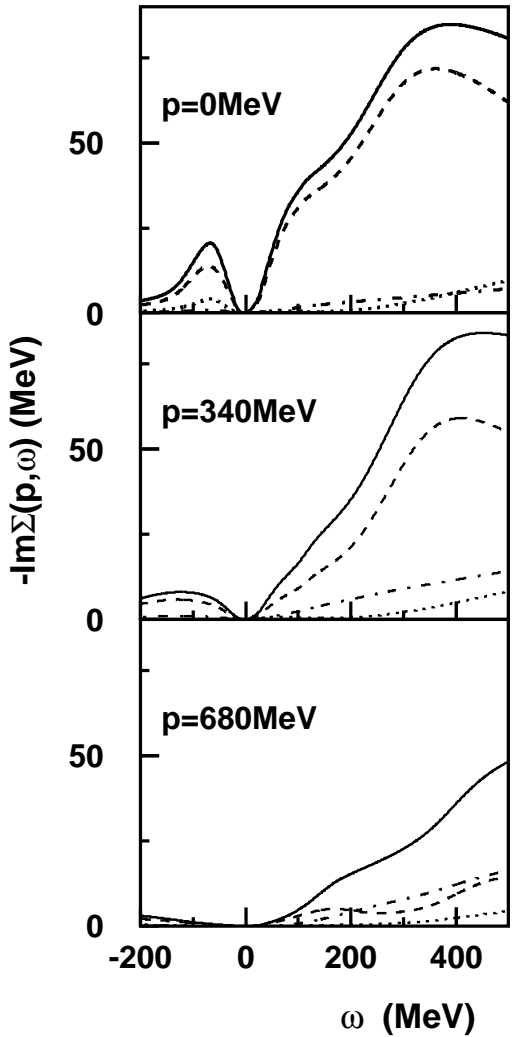}
\caption{\label{imagsd} The single particle width  $-\textrm{Im}\Sigma(p,\omega)$
 as function
  of energy for $p=0$, $340$,  and $680\textrm{MeV}$. The total width is
  plotted using the solid line. The dashed, dotted, and dashed-dotted lines
  represent the $^3S_1-^3D_1$, $^1S_0$, and $^3P_1$ 
partial waves contributions, respectively.}
\end{figure}

The large value of the scattering width  above the Fermi energy is due to
the short range part of the interaction potential. Splitting the total width
into contributions from different partial waves shows that the $^3S_1-^3D_1$
partial wave is dominant (Fig. \ref{imagsd}). For momenta up to twice the
Fermi momentum and energies around the Fermi energy, the deuteron partial wave
gives by far the most important contribution to the scattering width.
It is also in this kinematical region that the overall scattering width 
is the largest. We have observed a similar large contribution due to
this partial wave for the Monagan \cite{Mongan:1969dc} and Yamaguchi
\cite{yamaguchi} separable potentials. Consistently, for all these interactions
we obtain similar values for $Z_{p_F}\simeq.7$.
Large values of the scattering width at large energies lead to
 long tails in the
spectral functions.

The imaginary part of the self-energy at the quasi-particle pole is the
scattering width of the quasi-particle.
\begin{figure}
\includegraphics[width=0.48\textwidth]{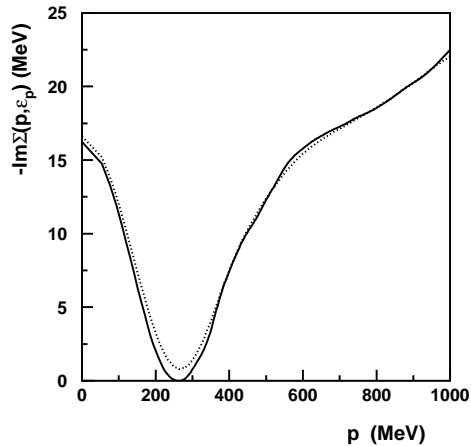}
\caption{\label{szerfig} The single particle width  $-\textrm{Im}\Sigma(p,\omega)$
at the
  quasiparticle pole as function of the momentum at $T=0$ 
(solid line) and  $T=4\textrm{MeV}$ 
(dotted line).}
\end{figure}
From Fig. \ref{szerfig} we see that around the Fermi momentum, narrow
quasi-particle excitation exist. The scattering width behaves as
$-\textrm{Im}\Sigma(p,\omega_p) \propto (p-p_F)^2$, as expected from 
restricted phase 
space for scattering \cite{abrikosov}.  The on-shell scattering width does not
reflect the very large values of the off-shell scattering width 
(Fig. \ref{imagfig}).  For momenta close to the Fermi surface it is small; and
 particles with large energies have  a scattering probability proportional
to the total density and cross section.
At finite temperature quasi-particles at the Fermi surface aquire a finite
life-time (Fig. \ref{szerfig}).


\section{Spectral function}

\begin{figure}
\includegraphics[width=0.48\textwidth]{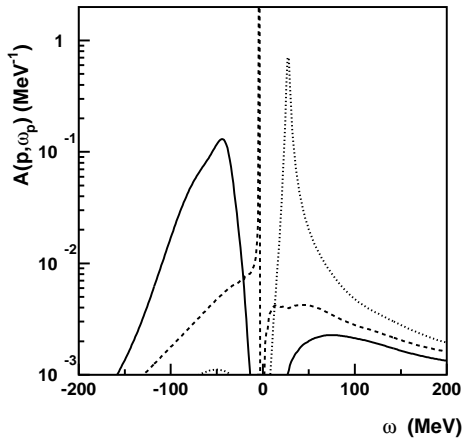}
\caption{\label{specfig}  The spectral function $A(p,\omega)$ as function of energy
  for $p=0$, $255$, and $350\textrm{MeV}$;  solid, dashed, and
  dotted lines, respectively.}
\end{figure}

The role of correlation induced by the medium can be judged by the
modifications of the spectral function. Nontrivial, dispersive self-energies
lead to broad spectral function with non-Lorentzian shapes and long tails.
High energy parts of the spectral functions could be revealed by electron
scattering experiments. It is important to calculate the form of the spectral
functions including contributions form short range correlations.
 In Fig. \ref{specfig} are presented the spectral functions for three
 representative momenta. For zero momentum the spectral function has a broad 
 peak below  and a background part above the Fermi energy. It is the reverse
 for $p=355\textrm{MeV}$. Following the behavior of the scattering width,
the spectral function for any momentum goes to zero at the Fermi energy.
For momenta close to the Fermi energy the spectral function has a very sharp
peak at the single-particle energy. Its has also a significant background part
which cannot be ignored in the sum rules or in the calculation of effective 
interactions between quasi-particles. Similarly as for the scattering width
the self-consistent calculation gives rise to a long tail in the spectral
function at negative energies.

The spectral function can be used to obtain the momentum distribution
\begin{equation}
n(p)=\int \frac{d\omega}{2\pi} A(p,\omega) f(\omega) \ .
\end{equation}
Short range correlations modifying the spectral function are reflected in the
nucleon momentum distribution. The free Fermi distribution is depleted below 
$p_F$ and a high momentum tail in $n(p)$ is formed.
\begin{figure}
\includegraphics[width=0.48\textwidth]{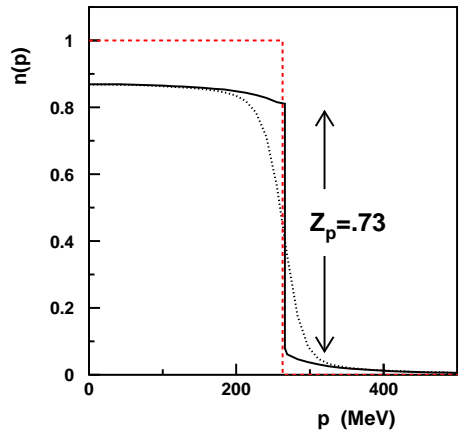}
\caption{\label{fermionfig} The momentum distribution of nucleons at zero
  temperature (solid line) and at $T=4\textrm{MeV}$ (dotted line).}
\end{figure}
At the Fermi momentum the discontinuity in the occupancy is reduced from $1$ to
$Z_{p_F}$ (\ref{zfac}).
The Fermi momentum in the interacting system should be the same as in the free
one \cite{Luttinger,Baym}. This relation is approximately fulfilled in the
present calculation, but not as well as  in our previous work using only $S$
wave interactions \cite{Bozek:2001tz}. A possible source of the discrepancy
could be the use of the  partial-wave expansion which spoils the 
$\Phi$-derivability of the self-energies. At finite temperature the 
Fermi surface is washed out, but the depletion of the Fermi sea and the
high momentum tail are the same. It means that the short range correlations
stay essentially the same at $T=4\textrm{MeV}$.
The same
 can also be seen by inspecting the self-energies. At finite temperature the
scattering width is increased only in the vicinity of the Fermi energy.

Long tails in the spectral function have  implications for saturation
properties of nuclear matter. Increased kinetic energy due to the high
momentum tail in $n(p)$ is compensated by increased removal energies due to
the negative energy tail of the spectral function.
 A detailed discussion at different densities will be
presented elsewhere.

\section{Summary}

We study the single-particle properties of nucleons
 in nuclear matter using a conserving,
in-medium $T$-matrix approximation.
The calculations are done in a self-consistent way with dressed propagators in
the ladder diagrams. To our knowledge, it is the first such calculation in the
literature using realistic interactions and several partial waves.
The  full discretization  of the spectral
function and self-energies allows to discuss the details of their 
energy dependence. We find that the basic features of a consistent
approximation to a Fermi liquid system are fulfilled. The scattering width is
zero at the Fermi  energy, and the quasi-particles at the Fermi surface have
infinite lifetime.  The momentum occupancy has a discontinuity of $Z_{p_F}$ at
the Fermi momentum.
The off-shell scattering width is very large at energies 
$\simeq 300-400\textrm{MeV}$ and momenta below $400\textrm{MeV}$. In this
region the main contribution comes from the $^3S_1-^3D_1$ partial wave.
The  $^3P_1$ and 
$^1S_0$  partial waves  give also important contributions 
to the on-shell self-energies.
 The scattering width on-shell is not very large, maximally
$26\textrm{MeV}$. Finite temperature effects are concentrated around the Fermi
surface. At finite temperature, the scattering width gets finite around 
the Fermi energy without
modifying the real single-particle potential and the short range correlations.
The quasi-particle renormalization factor is $Z_{p_F}\simeq.73$
and is largely independent on the interaction used. The effective mass is 
$m^\star\simeq.85m$.

\acknowledgments
This work was partly supported by the KBN
under Grant 2P03B02019.

\appendix

\section{\label{nummeth} Numerical methods}

Calculations  using self-consistent spectral functions require
the evaluation of energy integrals in Eqs. (\ref{teq}) and (\ref{imags}).
This is the main numerical difficulty for any self-consistent approximation;
 only in the last years self-consistent
 approaches have been applied to the study of  high $T_c$
superconductors \cite{Haussman1,Pedersen,kornilovitch,letz} and nuclear matter
\cite{di1,Dickhoff:1999yi,Bozek:1998su,Bozek:2001tz,Dewulf:2000jg}.
Some calculations are performed in the imaginary time formalism
\cite{Haussman1,Pedersen} which requires a numerical procedure 
for the analytical 
continuation to calculate the spectral function. A simpler way is to use the
real-time formalism  operating with the retarded $T$-matrix, the retarded
self-energy,
 and the spectral function \cite{kornilovitch,Bozek:1998su}. 
The real-time formalism
was also used for other calculations of the nuclear matter 
\cite{Dickhoff:1999yi,Dewulf:2000jg}
performed at zero temperatures. To deal with the off-shell propagation 
a numerical  parameterization of 
the energy dependence of the spectral function $A(p,\omega)$ by a set of
Gaussians has
 been used \cite{di1,Dickhoff:1999yi}. Alternatively, the spectral function can
be represented as a sum of $\delta$ functions \cite{letz,Dewulf:2000jg}.
The above 
two methods can be easily applied at zero temperature, where a narrow 
Gausian or one of the $\delta$ functions describes the quasi-particle peak for
momenta close to the Fermi surface.

\begin{figure}
\includegraphics[width=0.48\textwidth]{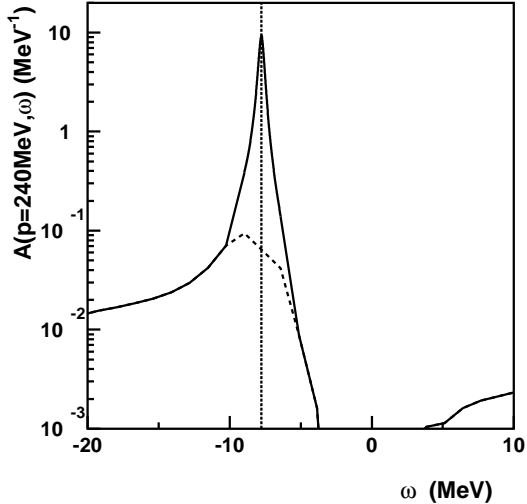}
\caption{\label{cutofffig} The spectral function $A(p,\omega)$ as function of
  the energy (solid line)  for $p=240\textrm{MeV}$
 and the corresponding smooth background part 
$B(p,\omega)$ (dashed line). The position of the 
 $\delta$ function 
representing  the quasi-particle peak at $\omega=-7.7\textrm{MeV}$ is 
indicated by the dotted line.}
\end{figure}

The alternative approach, here employed, uses a direct 
discretization of the spectral function and 
self-energies as functions of the energy. If the discretization in $\omega$ is
uniform the energy integrals in Eqs. (\ref{teq}) and (\ref{imags}) can be
performed efficiently by Fourier transforms. 
This method can be used directly 
at finite temperature \cite{Bozek:1998su,kornilovitch}, but not 
for low temperatures where the quasi-particles are very narrow around the 
Fermi surface (Fig. \ref{szerfig}). At zero temperature we have
\begin{equation}
A(p_F,\omega)=2\pi Z_{p_F} \delta(\omega-E_F) + B(p,\omega)
\end{equation}
where the background part $B$ is smooth and can be discretized.
Generally, for any momentum close to the Fermi momentum the
spectral function is rapidly changing in the vicinity of the quasi-particle 
peak; therefore, it is has to be split into a quasi-particle peak and a smooth 
background. The sharp peak is approximated by a $\delta$ function and the
smooth part is discretized.
We have found numerically  that  this separation in the spectral function
is needed if the width of the quasi-particle peak 
$-2\textrm{Im}\Sigma(p,\omega_p)$
is smaller than $\simeq 3$ times the spacing in the energy 
discretization $\Delta \omega$.
We use the following representation of the spectral function
\begin{equation}
\label{aspl}
A(p,\omega)=2\pi \mathcal{Z}_{p} \delta(\omega-\omega_p) + B(p,\omega) \ .
\end{equation}
The background part is defined as
\begin{equation}
B(p,\omega)=\left\{ \begin{array}{c} \frac{-2 \mathrm{Im}\Sigma(p,\omega)}
{(\omega-\omega_p)^2+\mathrm{Im}\Sigma(p,\omega)^2} \ ; \\ \text{for} \
(\omega-\omega_p)^2+\mathrm{Im}\Sigma(p,\omega)^2 > \eta \\ \\
{-2 \mathrm{Im}\Sigma(p,\omega)}/{
\eta} \ ; \\  \text{for}  \ 
(\omega-\omega_p)^2+\mathrm{Im}
\Sigma(p,\omega)^2 < \eta \end{array} \right. \ .
\end{equation}
The parameter $\eta$ is set to cut off the rapidly changing part of 
the spectral function. We use $\eta \simeq 4 \Delta \omega^2$.
The strength of the quasi-particle component $\mathcal{Z}_p$ 
in Eq. (\ref{aspl}) is 
adjusted to conserve the sum rule $\int \frac{d \omega}{2 \pi}
A(p,\omega)=1$. The weight of the singular part $\mathcal{Z}_p$ 
 is not the same as the renormalization factor 
$Z_p$ 
(Eq. \ref{zfac}). In Fig. \ref{cutofffig} is shown an example of the
separation of the background and singular parts of the spectral function 
$A(p=255\text{MeV},\omega)$.

For a separable interaction $V^\alpha(p,p^{'})
=\sum_{i, j}\lambda_{i,j}^\alpha g_i^\alpha(p)g_j^\alpha
(p^{'})$ the $T^\alpha$-matrix for the partial wave $\alpha=(JST)$ is $
\langle p|T^\alpha(\omega,P)|p^{'}\rangle
=\sum_{i, j} T^\alpha_{i j}(\omega, P)g_i^\alpha(p) 
g_j^\alpha(p^{'})$. Where the matrix $T^\alpha_{i j}= 
(\lambda_{i j}^\alpha -J_{i
  j}^\alpha)^{-1}$
with the coefficients $J_{i j}^\alpha$  given by the integrals
\begin{eqnarray}
J_{i j}^\alpha(\Omega,P)= \int \frac{d \omega_1 d\omega_2 d^3 k}{32 \pi^5}
g_i^\alpha(k)g_j^\alpha(k) \nonumber \\
\frac{A(p_1,\omega_1)A(p_2,\omega_2)}{\Omega-\omega_1-\omega_2+i\epsilon}
\left(1-f(\omega_1)-f(\omega_2) \right)  \ .
\end{eqnarray}
Substituting (\ref{aspl}) for the spectral functions we have
\begin{widetext}
\begin{eqnarray}
\label{imj}
{\rm Im}J_{i j}^\alpha(\Omega,P)=-
\int\frac{d \omega d^3k}{32 \pi^4} g_i^\alpha(k)g_j^\alpha(k)
B(p_1,\omega)\left(1-2f(\omega)\right)B(p_2,\Omega-\omega) & & \nonumber \\
-\int\frac{d^3k}{8\pi^3} g_i^\alpha(k)g_j^\alpha(k)
B(p_1,\Omega-\omega_{p_2}){\cal Z}_{p_2} \left(1-f(\Omega-\omega_{p_2})-
f(\omega_{p_2})\right) & & \nonumber \\
+{\rm Im} \int \frac{d^3k} {8\pi^3}
 g_i^\alpha(k)g_j^\alpha(k) \frac{ {\cal Z}_{p_1} {\cal Z}_{p_2}
\left(1-f(\omega_{p_1})-f(\omega_{p_2})\right)}
{\Omega -\omega_{p_1}-\omega_{p_2}+i\epsilon} \ .
\end{eqnarray}
\end{widetext}

The energy integral in the first term of Eq. (\ref{imj})
 is a convolution of the
functions $B(p_1,\omega)\left(1-2f(\omega)\right)$ and $B(p_2,\omega)$.
It can be evaluated by Fourier transformations using FFT transform algorithms
\cite{numericalrecipes}. Moreover, the Fourier transform and its inverse can
be performed outside of the momentum integral \cite{Bozek:1999rv}. The second
term of (\ref{imj}) is a standard two-dimensional integral without 
singularities. The integral in the last term of (\ref{imj}) is 
of the type often occurring in quasi-particle calculations, such as the
$G$-matrix or the quasi-particle $T$-matrix approximations. 
Angle averaging separately the
numerator and the denominator under the integral we have
\begin{widetext}
\begin{eqnarray}
{\rm Im} \int \frac{d^3k} {8\pi^3}
 g_i^\alpha(k)g_j^\alpha(k) \frac{ {\cal Z}_{p_1} {\cal Z}_{p_2}
\left(1-f(\omega_{p_1})-f(\omega_{p_2})\right)}
{\Omega -\omega_{p_1}-\omega_{p_2}+i\epsilon} & \simeq & \nonumber \\
{\rm Im} \int \frac{k^2d k} {2\pi^2}
 g_i^\alpha(k)g_j^\alpha(k) \frac{ <{\cal Z}_{p_1} {\cal Z}_{p_2}
\left(1-f(\omega_{p_1})-f(\omega_{p_2})\right)>_{P,k}}
{\Omega -<\omega_{p_1}+\omega_{p_2}>_{P,k}+i\epsilon} & =& \nonumber \\
-\frac{k_0^2}{2\pi}\frac{ <{\cal Z}_{p_1} {\cal Z}_{p_2}
\left(1-f(\omega_{p_1})-f(\omega_{p_2})\right)>_{P,k_0}}
{\partial<\omega_{p_1}+\omega_{p_2}>_{P,k}/\partial k|_{k=k_0}} & & 
\end{eqnarray}
\end{widetext}
where $k_0$ is the solution of
$$
\Omega=<\omega_{p_1}+\omega_{p_2}>_{P,k_0} \ .
$$
The angle average in the denominator $<\omega_{p_1}+\omega_{p_2}>_{P,k}$
is a function of the total and relative momentum ($P$ and $k$).
It can be represented using a one-dimensional function
\begin{equation}
<\omega_{p_1}+\omega_{p_2}>_{P,k}=\frac{2}{P k}\left(F(P/2+k)-F(P/2-k)\right) 
\end{equation}
with
$$F(x)=\int^{{x}}_0 pd p  \omega_{{p}} \ \ . $$ 
This effectively one-dimensional parameterization 
 allows to perform the integral corresponding to the
quasi-particle part of the spectral function very efficiently without using
any parabolic approximation for $\omega_p$. The same method can also
be used in  $G$-matrix calculations.
The real part of the integrals $J_{i j}^\alpha$ is obtained using 
the dispersion relation
\[
\textrm{Re} J_{i j}^{\alpha}(\Omega,P)=\int\frac{\Omega^{'}}{\pi}\frac{
\textrm{Im} J_{i j}^\alpha(\Omega^{'},P)}{\Omega^{'}-\Omega} \ .
\]

The calculation of the energy integral in the equation for the self-energy 
(\ref{imags}) proceeds very similarly.
We have 
\begin{eqnarray}
\label{imagsnum} & &
{\rm Im}
\Sigma(p,\omega) =\int\frac{d\omega_1d^3k}{16 \pi^4}
B(k,\omega_1) \nonumber \\ & &
 \langle({\bf p}-{\bf k})/2|{\rm Im}T({\bf p}
+{\bf k},\omega+\omega_1)|({\bf p}-{\bf k})/2\rangle_A \nonumber \\ & &
 \left( f(\omega_1)+b(\omega+\omega_1) \right) \nonumber \\
 & & + \int\frac{d^3k}{8 \pi^3} \nonumber \\ & &
\langle({\bf p}-{\bf k})/2|{\rm Im}T({\bf p}
+{\bf k},\omega+\omega_k)|({\bf p}-{\bf k})/2\rangle_A
\nonumber \\   & & \left( f(\omega_k)+b(\omega+\omega_k) \right) \ .
\end{eqnarray}
The  energy integral in the first term is again  of the form of a 
convolution and is calculated using Fourier transforms. The integral in the
second term is a standard two-dimensional integral.
Unlike in the $G$-matrix approximation, the calculation of the self-energy
requires the knowledge of the full off-shell $T$-matrix. 
Restricting oneself
to contributions from on-shell $T$-matrix leads to erroneous 
results \cite{schnellphd}.

In the iteration of the self-consistent set of equations (\ref{teq}, 
\ref{imags}, \ref{spectralf}, \ref{reals} and \ref{densitycon})
it is advantageous to  parameterize the energy dependence of the off-shell
quantities with respect to
the Fermi energy. The iterations are much faster, since a modification of
the chemical potential between iterations does not change the 
most important features of the $\omega$ dependence of $\Sigma(p,\omega)$
 and $T(\Omega,P)$, e.g. $\textrm{Im}\Sigma(p,\omega=0)=0$ and
the pairing singularity for $T(\Omega=0,P=0)$. 
 The absolute energy scale is recovered only in the
calculation of the total density (\ref{densitycon}) and final observables.
This way of proceeding with the iterations can make use of self-energies
calculated for other densities or temperatures to start a new iteration.

\section{N-N interaction and numerical parameters}

\label{NNint}

We use a separable parameterization of the Paris potential 
 \cite{parisseparable,parisseparable2}. It contains all the $J=0$, $J=1$,
$J=2$, and the $^3D_3-^3G_3$ partial waves.
For the most important $^1S_0$ and $^3S_1-^3D_1$ partial waves we
choose the rank $3$ and $4$ respectively. We have observed that the use of the
separable  parameterization of the Paris potential for the $^3P_1$ 
partial wave leads to very high values of the effective mass 
$m^\star\simeq.95m$. Since the $^3P_1$ phase shifts are not well reproduced 
by the parameterization of Ref. \cite{parisseparable}, 
we choose the Mongan $I$
parameterization \cite{Mongan:1969dc} for this partial wave. 
It is important to check
 that no unphysical bound states occur in the off-shell $T$-matrix for the
 given choice of $N-N$ potentials. 
The contribution from such unphysical
bound states, even far from the considered on-shell energies, would spoil
 the calculated self-energies.

The use of the Fourier transform algorithm for the calculation of energy 
integrals requires a fixed spacing in the energy grid. It means that we have
 to set finite ranges
for the kinematical variables in the model. The single particle momenta are in
the range $p<1700\textrm{MeV}$.
The total momentum of a nucleon pair is 
limited by $3400\textrm{MeV}$. The energy dependence of the $T$-matrix
$T(\Omega,P)$ is
calculated for $-2500\textrm{MeV}+P^2/4m < \Omega < 3500\textrm{MeV}$.
The energy range for the single particle self-energy is taken from
$-1700\textrm{MeV}$ to $3500\textrm{MeV}$.
 The energy dependent functions are discretized with a grid spacing of 
$1.28\textrm{MeV}$.

To obtain results with stability between iterations better than $1\%$
we need typically 7 iterations at zero temperature. The convergence of the
iterations is faster and more stable at finite temperature. 

\bibliography{../mojbib}

\end{document}